# High-Quality InSb Nanowires Grown by Molecular-Beam Epitaxy via Precise Catalyst Modulation

Wenlong Yang[1,#], Jiaye Xu[4,#], Shuai Yang[3], Wenyu Song[4], Haiyan Shi[1,2], Donghai Li[1,2], Xiyu Hou[1,2], Sijun Chen[3], Ke He[4,5,6], Hao Zhang[3,*], Dong Pan[1,2,*]

*[1]State Key Laboratory of Semiconductor Physics and Chip Technologies, Institute of Semiconductors, Chinese Academy of Sciences, Beijing 100083, China*

*[2]Center of Materials Science and Optoelectronics Engineering, University of Chinese Academy of Sciences, Beijing 100049, China*

*[3]College of Semiconductors, Southern University of Science and Technology, Shenzhen 518055, China*

*[4]State Key Laboratory of Low Dimensional Quantum Physics, Department of Physics, Tsinghua University, Beijing 100084, China*

*[5]Beijing Academy of Quantum Information Sciences, Beijing 100193, China*

*[6]Hefei National Laboratory, Hefei 230088, China*

[*]To whom correspondence should be addressed. E-mails: zhanghao@sustech.edu.cn; pandong@semi.ac.cn

[#]These authors contributed equally to this work.

**ABSTRACT:** Semiconductor InSb nanowires are a major material candidate for topological quantum computation. The nanowires in current devices were mostly grown using metal organic chemical vapor deposition, incompatible with *in situ* growth of superconductor, a crucial requirement for low-disorder devices. Growth of InSb nanowires using molecular-beam epitaxy (MBE) can address this issue. So far, studies on MBE-grown InSb remain limited due to an extremely narrow parameter window for the growth and reproducibility. Moreover, the use of foreign catalysts can introduce deep-level impurities. Here, we report a self-catalyzed MBE growth of InSb nanowires without those foreign catalysts. We first grow GaAs nanowire stems, and then convert Ga catalyst droplets into In droplets *in situ*. Axial heteroepitaxial growth of InSb can be realized through the In droplets on top of GaAs stems. We further find that a low flux suppresses radial growth, enabling linear length tuning from 0.8 to 2.6 μm with growth time, while diameter stays nearly constant. These InSb nanowires have a pure zinc-blende structure free of twins, stacking faults. Transport measurement reveals an electron mobility near 19 000 $cm^2/Vs$. Moreover, the observation of quasi-ballistic transport, quantum dots, and Kondo correlations demonstrates the coherent quantum transport capability of the MBE-grown InSb nanowires. Our work provides a substrate-independent, self-catalyzed route for the controllable synthesis of high-quality InSb nanowires, establishing a robust material platform for high-performance InSb-based electronic and quantum devices.

Among III-V semiconductors, InSb possesses the highest electron mobility and the narrowest bandgap, making it a highly attractive material for high-speed, low-power electronic devices as well as infrared optoelectronics [1-4]. In recent years, the small effective mass, strong spin-orbit coupling, and large Landé *g*-factor of InSb have fueled growing interest in epitaxially grown InSb nanowires for applications in detection and manipulation of Majorana fermions [5-14]. However, all these applications impose stringent demands on the crystal quality of InSb. Currently, two main approaches are employed for the growth of InSb nanowires. The first approach is direct growth on semiconductor substrates [15,16]. Because InSb possesses the largest lattice parameter among all III-V semiconductors, InSb nanowires grown directly on substrates often exhibit bending, twinning, polycrystalline formation, or severe stacking faults, which severely degrade the nanowire quality. To overcome the bottleneck associated with direct epitaxy, researchers have turned to axial heteroepitaxy of InSb nanowires on nanowire stems [8,17-24]. It should be noted that whether by direct growth or by source-switching processes, the current preparation of InSb nanowires still relies on foreign catalysts such as Au or Ag nanoparticles [8,11,12,15,17-19,20-26]. The introduction of foreign catalysts, in particular, can induce deep-level impurities, deteriorate carrier transport properties, and is incompatible with complementary metal-oxide-semiconductor (CMOS) technology [25,26]. In contrast, self-catalyzed growth, which uses group-III elements as the catalyst, fundamentally avoids deep-level impurities and offers significant advantages in process compatibility [27-29].

Nevertheless, relevant studies on self-catalyzed InSb nanowire growth remain very limited [30-32]. In particular, reported self-catalyzed InSb nanowires either exhibit crystalline defects or suffer from an extremely narrow growth window. Therefore, there is an urgent need to develop a novel self-catalyzed technique for the growth of high-quality InSb nanowires.

On the device side, nearly all InSb nanowires used in Majorana-related research were grown using metal organic chemical vapor deposition (MOCVD) [7-12]. However, MOCVD equipment does not permit the *in situ* low-temperature epitaxy of superconductors. Consequently, the unavoidable disorder at the semiconductor-superconductor interface—which gives rise to soft superconducting gaps and trivial Andreev bound states—has long plagued transport measurements in such hybrid systems. Molecular-beam epitaxy (MBE) could in principle address this long-standing obstacle, as superconductors could be epitaxially deposited *in situ* at low temperature. The MBE approach is, however, extremely challenging due to the fact that Sb acts as a surfactant; under ultra-high vacuum conditions, Sb strongly suppresses axial growth. It is therefore imperative to develop a wholly new MBE approach for InSb nanowire growth, enabling the subsequent low-temperature *in situ* epitaxy of superconductors and providing a critical material platform for topological quantum computation and related research.

To address the above challenges, we report a new self-catalyzed growth scheme for InSb nanowires on Si (111) substrates using MBE. We propose and experimentally validate a group-III catalyst modulation strategy. Specifically, we employ the well-established self-catalyzed GaAs nanowire system as both a growth template and a catalyst carrier. By modulating the catalyst composition, the Ga catalyst droplet is converted into an In droplet *in situ*, thereby enabling axial heteroepitaxial self-catalyzed vapor-liquid-solid (VLS) growth of InSb on top of GaAs. We find that, at the same Sb/In beam equivalent pressure ratio (BEP), the radial growth rate of InSb nanowires can be modulated by employing different beam fluxes, and that high-aspect-ratio InSb nanowires can be obtained by adjusting the growth time accordingly. Microstructural characterization and compositional analysis confirm that the as-grown InSb nanowires exhibit a pure zinc-blende structure, free from twins and stacking faults. Quantum transport characterizations on these InSb nanowires reveal decent electron mobility, quasi-ballistic transport, quantum dot and Kondo effect, establishing these as key signatures of the coherent quantum state. Our work provides a general strategy for *in situ* replacement of group-III catalyst elements at the nanowire tip, which is substrate-independent and can be extended to the self-catalyzed, controllable growth of high-quality III-V semiconductor nanowires on arbitrary substrates. In particular, the successful MBE growth of high-quality self-catalyzed InSb nanowires provides a robust basis for both the low-temperature *in situ* epitaxy of high-quality InSb-superconductor heterojunction nanowires and the ensuing investigations into topological quantum computation.

## EXPERIMENTAL DETAILS

**Nanowire growth.** All InSb nanowires were grown in an MBE system (VG V80H) using commercial n-type Si (111) wafers as substrates. Prior to growth, the Si substrates were subjected to the following pre-treatment: the substrates were immersed in an HF solution (8%) for approximately 40 seconds to remove the native oxide layer. Subsequently, they were dipped into a mixture of $H_2SO_4$ (98%) and $H_2O_2$ (30%) at a volume ratio of 4:1 for 15 seconds to form a fresh thin oxide layer on the substrate surface.

The GaAs nanowires were grown as follows. First, the Ga source shutter was opened for 2 minutes to ensure the formation of initial Ga catalyst droplets. Then, the As source shutter was opened, and GaAs nanowires were grown for 25 minutes at a V/III ratio of 20 and a growth temperature of 550 °C. After growth, the Ga source shutter was closed, while the As flux was maintained for an additional 5 minutes before being turned off. Subsequently, the temperature was lowered to 400 °C, the In flux was started for 1 minute, followed by the Sb flux, and InSb nanowires were grown at a V/III ratio of 1.5.

**Microstructural analysis.** The morphology of the nanowire samples was characterized by scanning electron microscopy (SEM, FEI Nano SEM 650, operated at 10 kV). The crystal structure of the nanowires was investigated by transmission electron microscopy (TEM, JEM F200, operated at 200 kV). Elemental distribution in the nanowires was studied by energy-dispersive X-ray spectroscopy (EDS) in scanning

TEM mode, using the JEM F200 TEM system. TEM analysis samples were prepared as follows: the samples were placed in ethanol and ultrasonically treated to detach the nanowires from the growth substrate; subsequently, the suspension was drop-cast onto a carbon-supported film. For the preparation of cross-sectional TEM samples of the nanowires, a Pt protective layer was first deposited on the target area of the sample. Then, ion beam cutting and extraction were performed, and the thin lamella was transferred onto a specialized TEM sample holder. Finally, low-voltage ion beam milling and cleaning were carried out to obtain electron-transparent thin lamellae suitable for TEM analysis.

**Electrical characterization.** The InSb nanowires were then transferred from the growth chip to a $SiO_2$/Si substrate by wipes of clean room tissues. The Si is p-doped, acting as a back gate. The 300-nm-thick $SiO_2$ is the dielectric layer. Ti/Au metal contacts were deposited using standard electron beam lithography. Before the contact deposition, a short (120 s) Ar treatment was performed in situ to remove surface oxides for ohmic contacts. Two types of devices were fabricated based on the contact spacing (channel length). Mobility devices have a long channel length (~ 1 μm). Quantum point contact (QPC) devices and quantum dot devices have a relatively short channel (150-300 nm). Both devices were measured in a dilution refrigerator at a base temperature below 50 mK. The measurement set-up was standard two-terminal.

**Catalyst-Mediated Growth of InSb Nanowires.** Figure 1 illustrates the complete experimental procedure for self-catalyzed growth of InSb nanowires on Si (111) substrates using the proposed group-III catalyst modulation method, which consists of six key steps. First, Ga catalyst droplets are deposited on the Si substrate to serve as catalyst precursors for the subsequent self-catalyzed growth of GaAs nanowires (Figure 1a). Subsequently, As flux is introduced into the chamber. When the Ga and As in the droplets become supersaturated, GaAs nanowires nucleate and precipitate at the droplet-substrate interface, forming GaAs nanowires perpendicular to the substrate (Figure 1b). The inset in Figure 1b is an SEM image of a GaAs nanowire, on which a Ga droplet is clearly observed at the tip. The next critical steps for the axial heteroepitaxy of InSb nanowires on GaAs nanowires are the partial consumption of the Ga droplets and the replacement/formation of In droplets (Figure 1c-d). After GaAs nanowire growth, the Ga flux is terminated while the As flux is maintained for a certain period at the same temperature. The relatively high temperature causes the Ga droplets to evaporate or react with As to form GaAs, thereby removing them. The inset in Figure 1c shows an SEM image of a GaAs nanowire, where the Ga droplet is partially consumed. Figure S1 shows TEM images of the catalyst state on different GaAs nanowires after As treatment, demonstrating that the residual state of the Ga catalyst droplets on top of the GaAs nanowires can be precisely controlled by adjusting the As treatment time.

During the gradual elimination of Ga droplets, the As flux was terminated, and the In flux was introduced upon ramping the temperature to the InSb growth conditions. In the absence of As supply, In did not immediately crystallize but instead accumulated on the surface, forming metallic droplets. By adjusting the In flux intensity or the exposure duration, the droplet size could be directly controlled. Residual Ga droplets served as a pre-existing liquid-phase template to facilitate the aggregation of In atoms; these Ga droplets effectively trapped nearby adatoms [33], substantially reducing the energy barrier for the independent nucleation of In on the GaAs top facet and ensuring the morphological stability of the catalyst droplets during the replacement process. With sustained In flux supply, the Ga within the alloy droplets was progressively diluted and replaced by In, ultimately yielding catalyst droplets predominantly composed of In. By extending the In flux duration, the size of the In-based catalyst droplets could be incrementally increased, thereby establishing conditions conducive to self-catalyzed InSb nanowire growth at the droplet apex (Figure 1e). Figure S2 presents the elemental distribution maps of the GaAs nanowires after the catalyst regulation procedure, confirming that the dominant constituent at the nanowire tips had been successfully switched from Ga to In droplets. Once the In droplets were fully formed, the Sb flux was introduced to initiate the growth of InSb nanowires (Figure 1f). The inset in Figure 1f presents an SEM image of a typical InSb nanowire obtained using this strategy. The underlying mechanism of the transformation of Ga droplets into In droplets is shown in S3 in the Supplemental Material.

Figure 2 presents a comparison of SEM morphologies of nanowires grown under different beam fluxes and growth times at the same V/III BEP ratio. Figure 2a and Figure 2b show the samples grown for 25 and 60 minutes, respectively, under a V/III

BEP ratio of 1.58 with high beam fluxes (In BEP = $0.97\times10^{-6}$ mbar, Sb BEP = $1.53\times10^{-6}$ mbar). At a growth time of 25 minutes, the nanowires have a length of approximately 0.8 μm and a diameter of about 250 nm. When the growth time is extended to 60 minutes, the length increases significantly to about 2.4 μm, while the diameter also markedly increases to approximately 400 nm. However, under the same V/III ratio but with reduced beam fluxes (In BEP = $1.53\times10^{-7}$ mbar, Sb BEP = $2.43\times10^{-7}$ mbar), as the growth time is prolonged from 25 to 60 minutes, the nanowire length also increases from about 0.8 μm to a maximum of about 2.6 μm, whereas the diameter undergoes almost no significant change. The observed diameter increase under high total beam flux is attributable to the enhanced supersaturation of In within the catalyst droplet and the resulting sidewall deposition. In self-catalyzed VLS growth, the In droplet functions as both a catalyst and a reactant reservoir. When the total beam flux is elevated, the incident In flux onto the substrate and nanowire surfaces increases substantially [34]. Given a constant V/III ratio, the Sb flux rises correspondingly. Under such high-flux conditions, the accumulation rate of In within the droplet may exceed its incorporation rate into the solid InSb lattice at the crystallization interface, leading to droplet volume expansion and a consequent change in contact angle. More critically, the high flux induces a sharp increase in the concentration of In adatoms adsorbed on the nanowire sidewalls. According to surface diffusion theory, the surface diffusion length of adatoms depends on their residence time and surface mobility. However, under high flux, sidewall adatom coverage rapidly reaches saturation; consequently, a large number of In adatoms cannot migrate to the top droplet in time to contribute to axial growth. Instead, they directly react with incident Sb atoms on the sidewalls, giving rise to non-catalytic VS deposition [35]. In contrast, at lower beam flux, the number of In adatoms on the sidewalls is considerably smaller. Moreover, as the sidewall surface area relative to the top droplet cross-section changes during growth, the limited adatoms preferentially migrate via long-range surface diffusion toward the droplet-solid interface at the apex. Since the sidewalls lack sufficient adatoms for non-catalytic nucleation, radial growth is effectively suppressed, and the diameter thus remains near its initial value of approximately 150 nm.

**Microstructure and Elemental Distribution of InSb Nanowires.** Figure 3a presents a low-magnification TEM image of a single representative InSb/GaAs nanowire, clearly showing a two-segment structure along the axial direction: a bottom GaAs nanowire segment and a top InSb nanowire segment. The two segments are sharply delineated, and the overall morphology is straight with a uniform diameter, indicating that after catalyst replacement, the axial growth of the InSb nanowire segment maintains good growth stability. Figure 3b-d show high-resolution TEM images of the InSb segment, the InSb/GaAs heterointerface region, and the GaAs segment, respectively. The InSb segment (Figure 3b) exhibits highly ordered lattice fringes free of stacking faults and twin defects, indicative of a pure zinc-blende structure. The GaAs segment (Figure 3d) similarly displays single-crystalline features with continuous and intact lattice fringes. At the InSb/GaAs interface region (Figure 3c), the lattice fringes of the two materials directly connect; no misfit dislocations, stacking faults, twins, or amorphous transition layers are observed at the interface. High-resolution TEM characterization confirms that the axial heteroepitaxy from the GaAs stem to the InSb top segment achieves a crystalline quality transition with extremely low defect density. Figure 3e shows a high-angle annular dark-field scanning transmission electron microscopy (HAADF-STEM) image of the tip of a single InSb nanowire. A spherical catalyst particle at the nanowire tip can be clearly observed. Figure 3f presents the corresponding EDS elemental mapping of the same region, revealing that In and Sb are uniformly distributed across the radial cross-section of the nanowire. The elemental map of the tip catalyst particle shows a dominant In signal, corresponding to an In-Sb alloy catalyst. Figure 3g presents an EDS line scan acquired along the nanowire axis across the catalyst region; the quantitative results further confirm that the catalyst particle is composed predominantly of In, thereby demonstrating that the InSb nanowire grows via a self-catalyzed VLS process. Figure 3h shows a HAADF-STEM image of the InSb/GaAs interface region. Since the contrast in HAADF imaging is approximately proportional to the square of the atomic number, the InSb region appears brighter while the GaAs region appears darker, forming a sharp and clear contrast boundary at the interface. Figure 3i presents the EDS elemental mapping of the interface region. Ga and As signals are concentrated on the GaAs side of the interface, while In and Sb signals are concentrated on the InSb side, with no obvious intermixing across the interface. Figure 3j shows an EDS line scan along the nanowire axial direction across the interface, quantitatively illustrating the spatial evolution of elemental intensities: the Ga and As signals drop sharply to near zero at the interface, while the In and Sb signals rise abruptly and quickly reach stable values, indicating no significant elemental interdiffusion between the InSb and GaAs segments (The Ga content in the InSb nanowires is shown in S4 in the Supplemental Material).

To further verify the crystalline integrity of the InSb nanowires, cross-sectional lamellae were prepared perpendicular to the nanowire axis using focused ion beam (FIB) milling, and atomic-scale characterization was performed by aberration-corrected HAADF-STEM. Figure 4a shows an overview HAADF-STEM image of the InSb nanowire cross-section, which exhibits a near-hexagonal geometry. Over the entire cross-sectional area, the contrast is uniform and continuous, with no observable defects. Figure 4b-i present high-resolution HAADF-STEM atomic images taken from different regions of the cross-section. Under high magnification, each region displays highly ordered atomic arrangements. In the atomic images of Figure 4b-i, the In and Sb atoms show consistent atomic ordering from the center to the periphery, with no structural differences between the interior and near-surface regions, nor any lattice distortion caused by localized strain fields. The EDS elemental maps of the cross-section (Figure S4) further show a homogeneous distribution of In and Sb elements across the cross-section. These results confirm the high-quality fabrication of self-catalyzed InSb nanowires in this study, providing a material foundation for the subsequent construction of electronic and quantum devices based on these InSb nanowires. It is worth noting that, owing to the use of $As_4$ in our MBE system, the As background pressure cannot be rapidly eliminated after the growth of GaAs nanowires. Consequently, a certain distribution of As is present in the subsequently grown InSb nanowires (see Figure S3). However, point-resolved EDS analysis indicates that the As content within the InSb nanowires is very low (~ 4%). In our future studies, this As background issue is expected to be resolved by employing a cracker source for As (providing $As_2$) instead of $As_4$.

**Electrical Properties of InSb Nanowires.** We next investigate the quantum transport of these InSb nanowires. Figure 5a shows the SEM of a mobility device. The channel length is around 1 μm. Its pinch-off curve, i.e. the conductance $G$ as a function of back gate voltage $V_{BG}$, is shown in Figure 5b. The gate sweeping direction is from negative to positive (black arrow). The bias voltage $V$ was fixed at 10 mV, and the current $I$ was measured. $G = 1/(V/I - R_{filter})$, where $R_{filter}$ is the series resistance (~ 3.5 kOhm) contributed by fridge filters. We then fit this pinch-off curve using the formula: $G = (R_s + L^2/(\mu C(V_{BG} - V_{th})))^{-1}$, where $L$ is the channel length, $R_s$ is the contact resistance, $V_{th}$ is the threshold voltage, $\mu$ is the mobility and $C$ is the capacitance between the channel and the back gate. To extract the mobility, one has to estimate the capacitance first. Figure 5c shows the cross-sectional STEM of this device after its measurement. The back gate (p-doped Si) and $SiO_2$ can be clearly identified. Based on Figure 5c, we estimate $C = 2.44 \times 10^{-17}$ F using a finite element method following Ref. [36]. The mobility fit (red curve in Figure 5b) then yields $\mu$ = 19, 000 $cm^2/Vs$. This value is not the highest but decent. For further improvement, one can optimize the growth condition to remove the residue As in the InSb nanowire. Another direction is to cap the InSb surface to prevent it from oxidization. The EDS map in Figure 5d highlights the surface oxides of the nanowire, which can be a major disorder source.

We also explored a short channel device as shown in Figure 5e. The channel length is around 320 nm. Figure 5f shows its differential conductance ($G = dI/dV$) as a function of $V_{BG}$ and magnetic field ($B$). The $B$ direction is in-plane and nearly perpendicular to the nanowire, as shown in Figure 5e. As $B$ increases, plateau features develop, suggesting the ballistic regime and suppression of back scattering. We labeled the $e^2/h$ plateau as "0.5" in the color map. The lower panel shows a line cut at 4.15 T, where the 0.5, 1.0, and 1.5 plateaus (in units of $2e^2/h$) can be revealed. A contact resistance of 1 kΩ is assumed and subtracted from this device. Based on the size of this 0.5 plateau and its evolution in $B$, we extracted a Landé g-factor of ~ 25.

For additional data on mobility and ballistic transport, we refer to S6 in the Supplemental Material. In Figure S5, we also show a short channel device where quantum dot forms, resulting in Kondo effects.

In summary, we have developed a self-catalyzed MBE approach for the growth of high-quality InSb nanowires on Si (111) substrates via a precise group-III catalyst modulation strategy. By using self-catalyzed GaAs nanowires as growth templates and converting Ga catalyst droplets into In droplets *in situ*, we successfully achieve axial heteroepitaxial VLS growth of InSb on GaAs stems. Systematic growth studies reveal that, at a fixed Sb/In BEP ratio, the beam flux critically governs the nanowire morphology, with low fluxes effectively suppressing radial growth and enabling linear length tuning while maintaining a nearly constant diameter. Microstructural characterization and elemental analysis demonstrate that the InSb segments possess a pure zinc-blende structure free of twins and stacking faults, with atomically sharp GaAs/InSb heterointerfaces. Electrical transport measurements show an electron mobility of 19, 000 $cm^2/Vs$. Ballistic transport can be achieved in a short-channel device (~320 nm long) at finite magnetic fields, evidenced by the observation of quantized plateaus. This substrate-independent, self-catalyzed route not only provides a robust material platform for InSb-based high-speed electronics and topological quantum devices, but also offers a general strategy for the controllable epitaxy of other high-quality III-V semiconductor nanowires on arbitrary substrates.

## ASSOCIATED CONTENT

### Supporting Information

TEM images of Ga droplets on GaAs nanowire stems under different As flux treatment durations. TEM images and EDS elemental distribution maps of GaAs nanowires after the catalyst regulation procedure. The underlying mechanism of the transformation of Ga droplets into In droplets. The Ga content in the InSb nanowires. TEM image and EDS elemental distribution maps acquired from the cross-sectional views of the InSb nanowires. Additional data on mobility and ballistic transport.

### Author Contributions

[#](W. Yang, J. Xu) These authors contributed equally to this work.

### Notes

The authors declare no competing financial interest.

### Acknowledgements

This work was supported by the National Natural Science Foundation of China (Grants 12374459, 61974138 and 92065106), the Quantum Science and Technology-National Science and Technology Major Project (Grant 2021ZD0302400), the National Key Research and Development Program of China (Grant 2025YFA1411400), and the Strategic Priority Research Program of Chinese Academy of Sciences (Grant

XDB0460000). D. P. also acknowledges the support from the State Key Laboratory of Micro-nano Engineering Science (Grant MES202601) and Youth Innovation Promotion Association, Chinese Academy of Sciences (Nos. 2017156 and Y2021043).

## Figures and figure captions

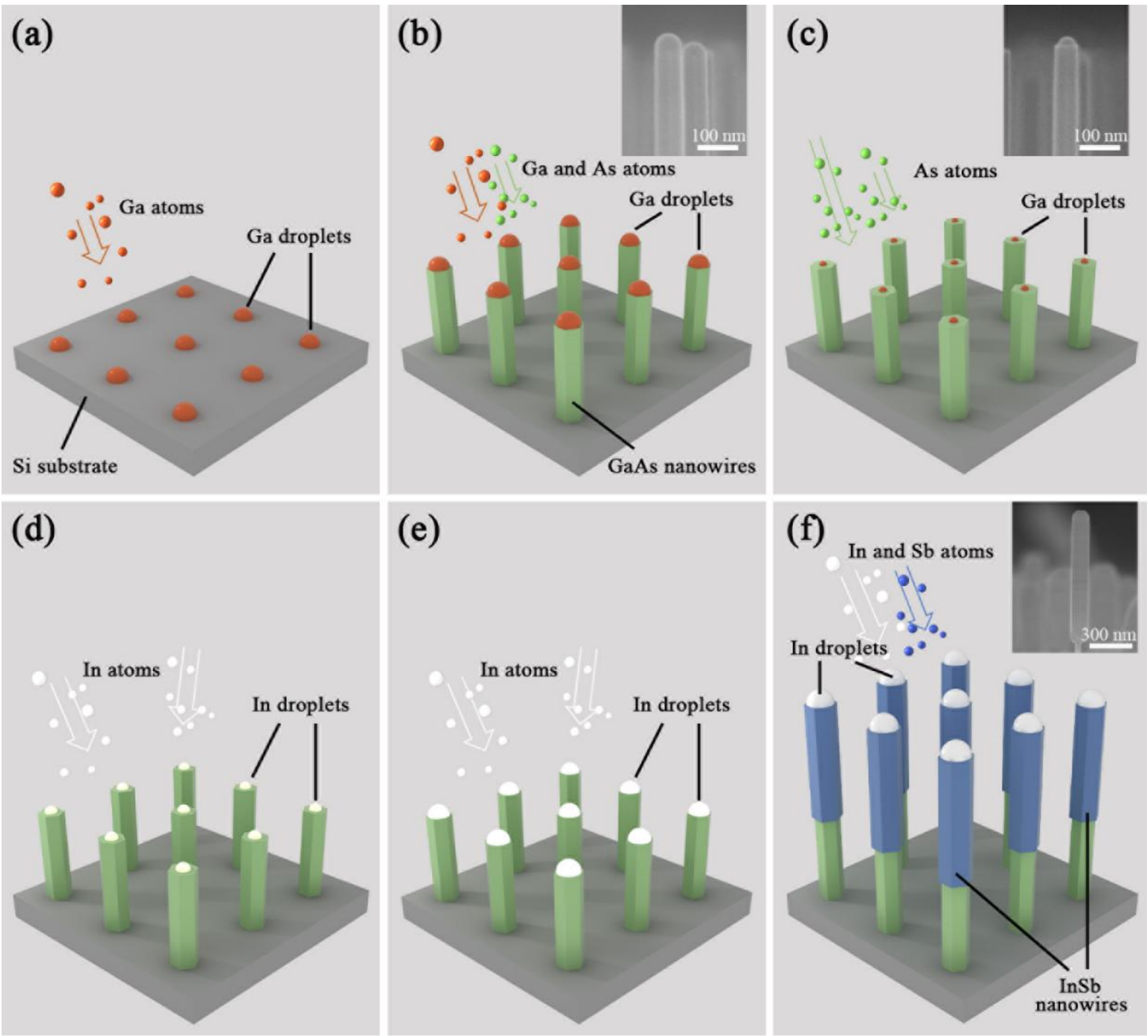

**Figure 1.** Schematic illustration of the growth process of high-quality self-catalyzed InSb nanowires on Si (111) substrates via precise catalyst modulation. (a) Deposition of Ga catalyst droplets on the Si substrate. (b) Introduction of As flux to initiate GaAs nanowire growth. (c) Maintaining As flux to consume the Ga catalyst droplets. (d) Introduction of In flux to start the accumulation of In catalyst droplets on the GaAs nanowires, which continues to a certain amount (e). (f) Introduction of Sb flux to initiate InSb nanowire growth. The insets in (b), (c), and (f) are typical SEM images of the nanowire samples.

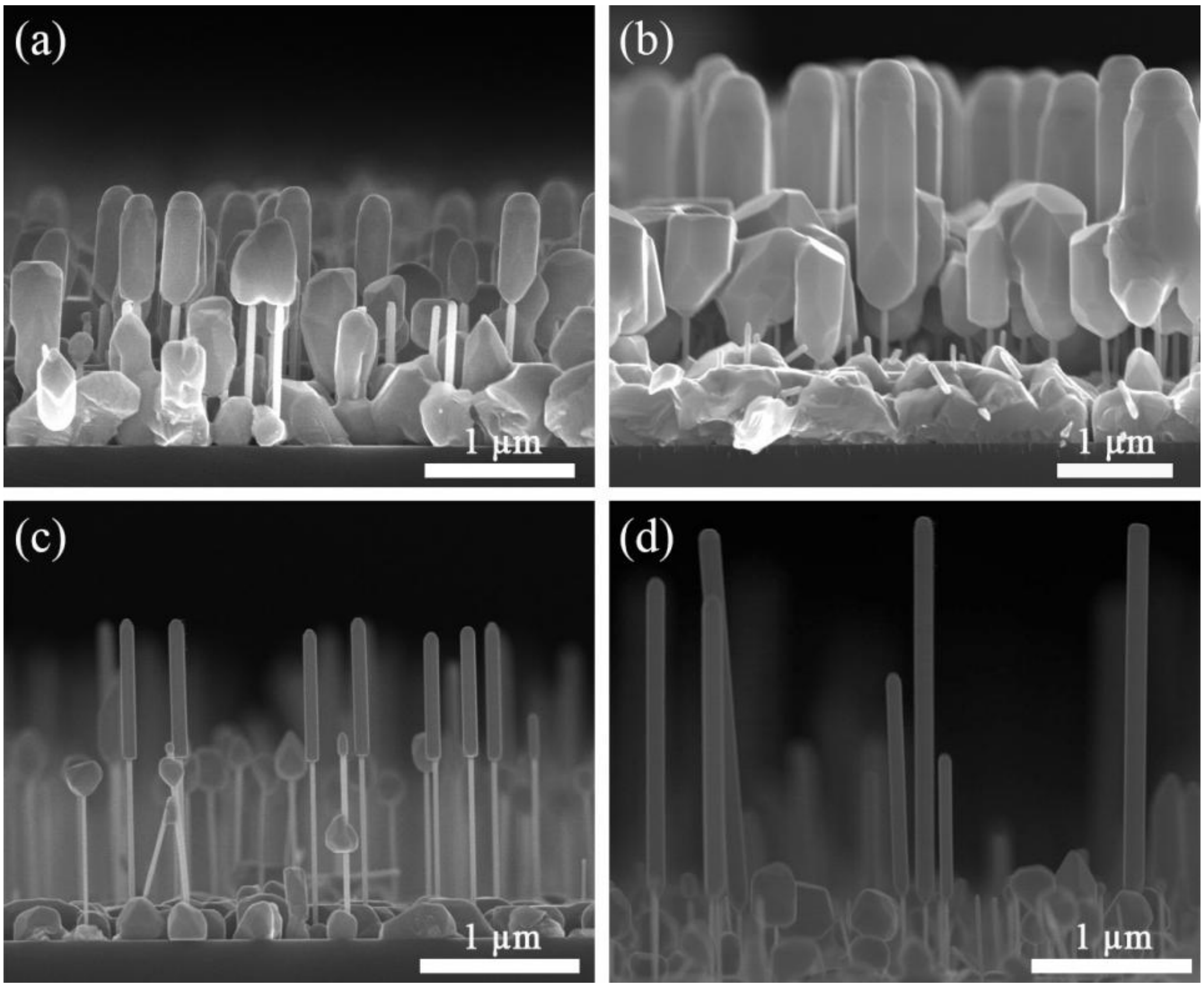

**Figure 2.** Cross-sectional SEM images of nanowire samples grown under the same Sb/In ratio but with different beam fluxes and growth times. (a) Sample grown for 25 min at In BEP = $0.97\times10^{-6}$ mbar and Sb BEP = $1.53\times10^{-6}$ mbar. (b) Sample grown for 60 min under the same high-flux conditions. (c) Sample grown for 25 min at In BEP = $1.53\times10^{-7}$ mbar and Sb BEP = $2.43\times10^{-7}$ mbar. (d) Sample grown for 60 min under the same low-flux conditions. All GaAs nanowires were grown for 25 minutes at a V/III ratio of 20 and a growth temperature of 550 °C.

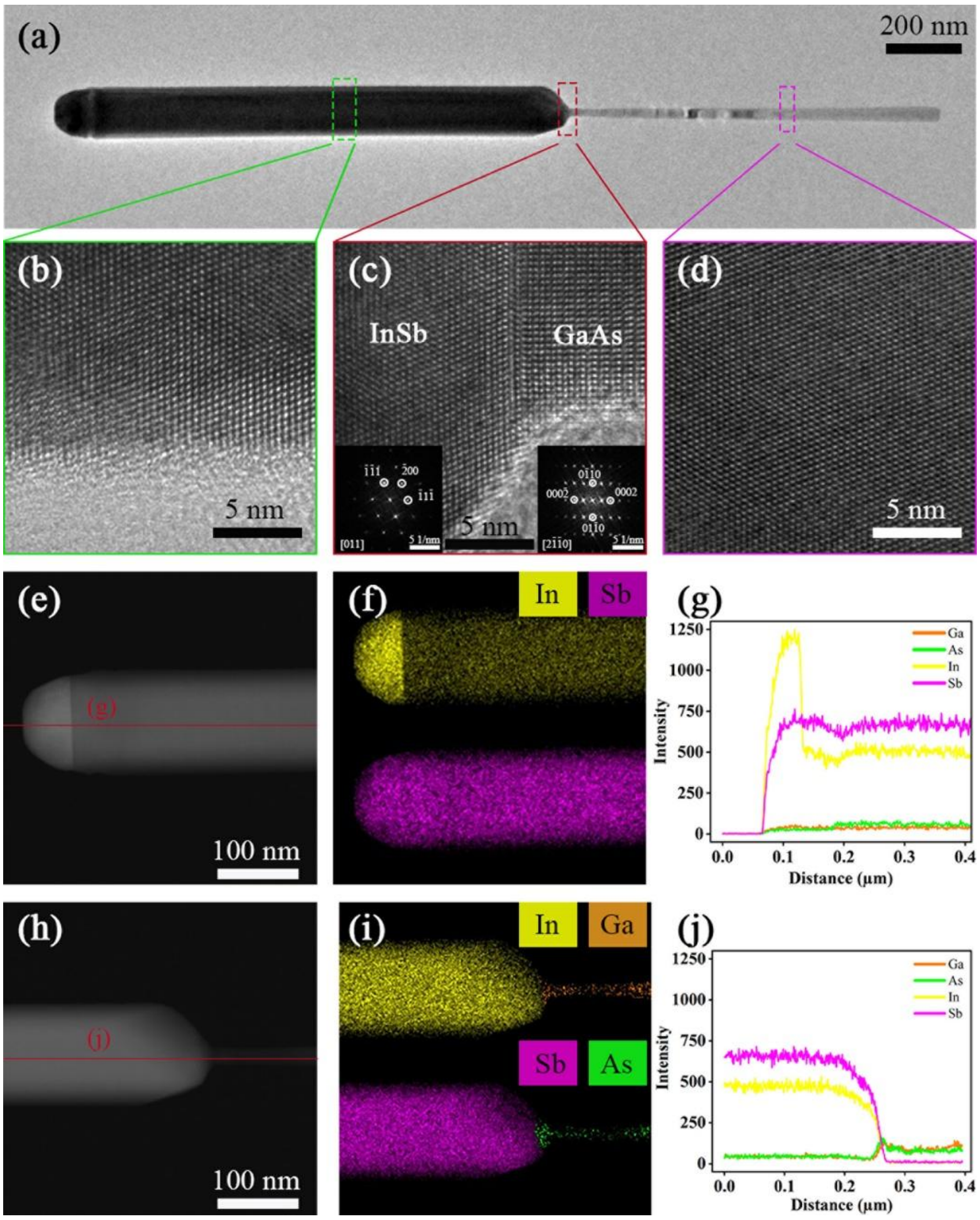

**Figure 3.** Microstructure and elemental distribution of InSb nanowires self-catalytically grown along the axial direction of GaAs nanowires. (a) Low-magnification TEM image of a single representative GaAs/InSb nanowire. (b-d) High-resolution TEM images of the InSb segment, the InSb/GaAs interface region, and the GaAs segment, corresponding to the green, red, and purple boxes in (a), respectively. (e) HAADF-STEM image of the tip region of an InSb nanowire. (f) EDS elemental mapping of the region shown in (e). (g) EDS line scan intensity profile along the nanowire axial direction across the catalyst region in (e). (h) HAADF-STEM image of an InSb nanowire region. (i) EDS elemental mapping of the region shown in (h). (j) EDS line scan intensity profile along the nanowire axial direction across the catalyst region in (h). The insets in the lower-left and upper-right corners of (c) are the fast Fourier transform patterns derived from the high-resolution TEM images of InSb and GaAs, respectively. The scale bars in the insets in (c) are 5 1/nm.

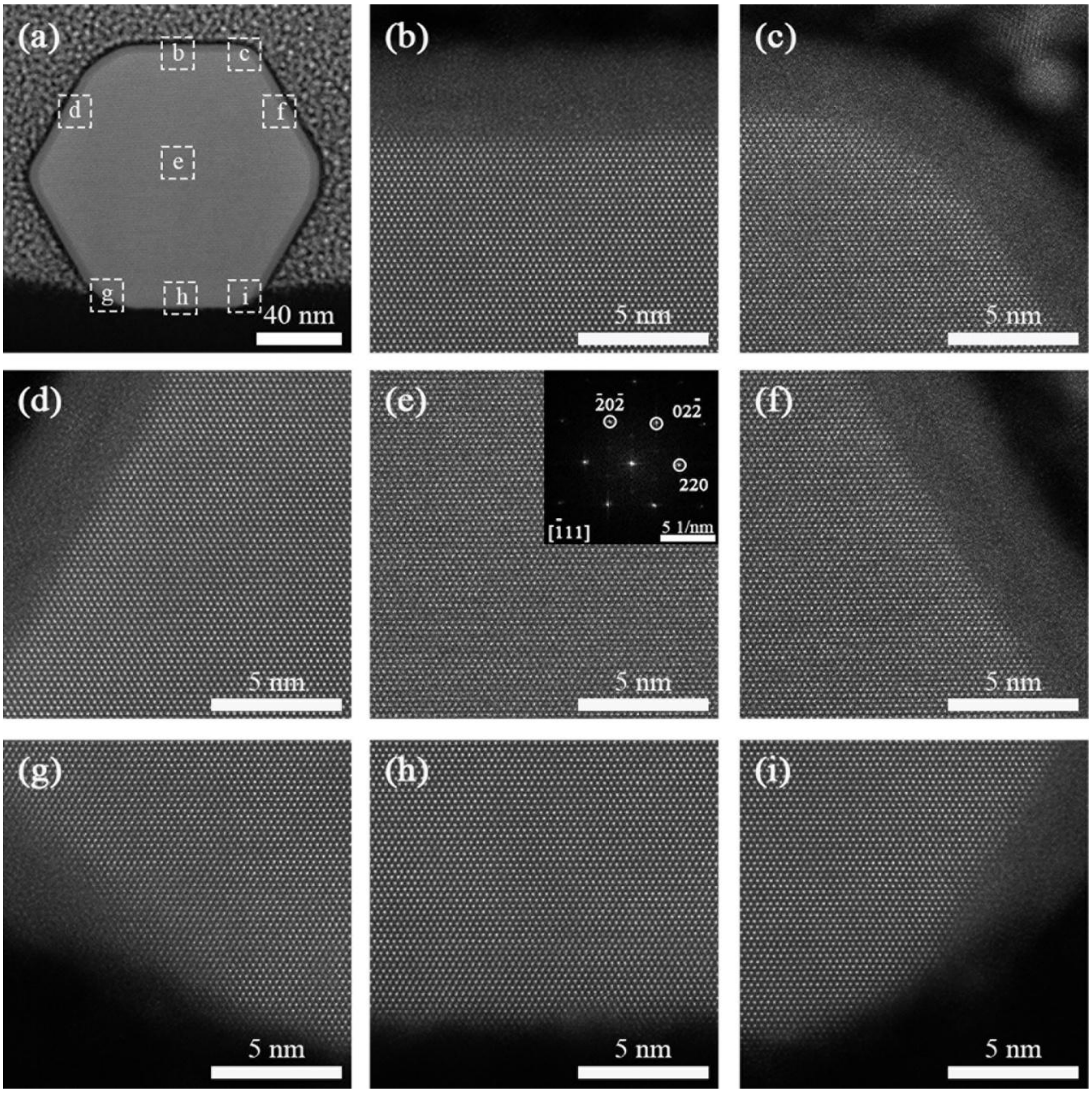

**Figure 4.** Cross-sectional microstructure of InSb nanowires self-catalytically grown along the axial direction of GaAs nanowires. (a) Low-magnification HAADF-STEM overview image of the cross-section of an InSb nanowire; (b-i) High-resolution HAADF-STEM images of different cross-sectional regions marked in (a). The inset of panel (e) presents the corresponding fast Fourier transform pattern.

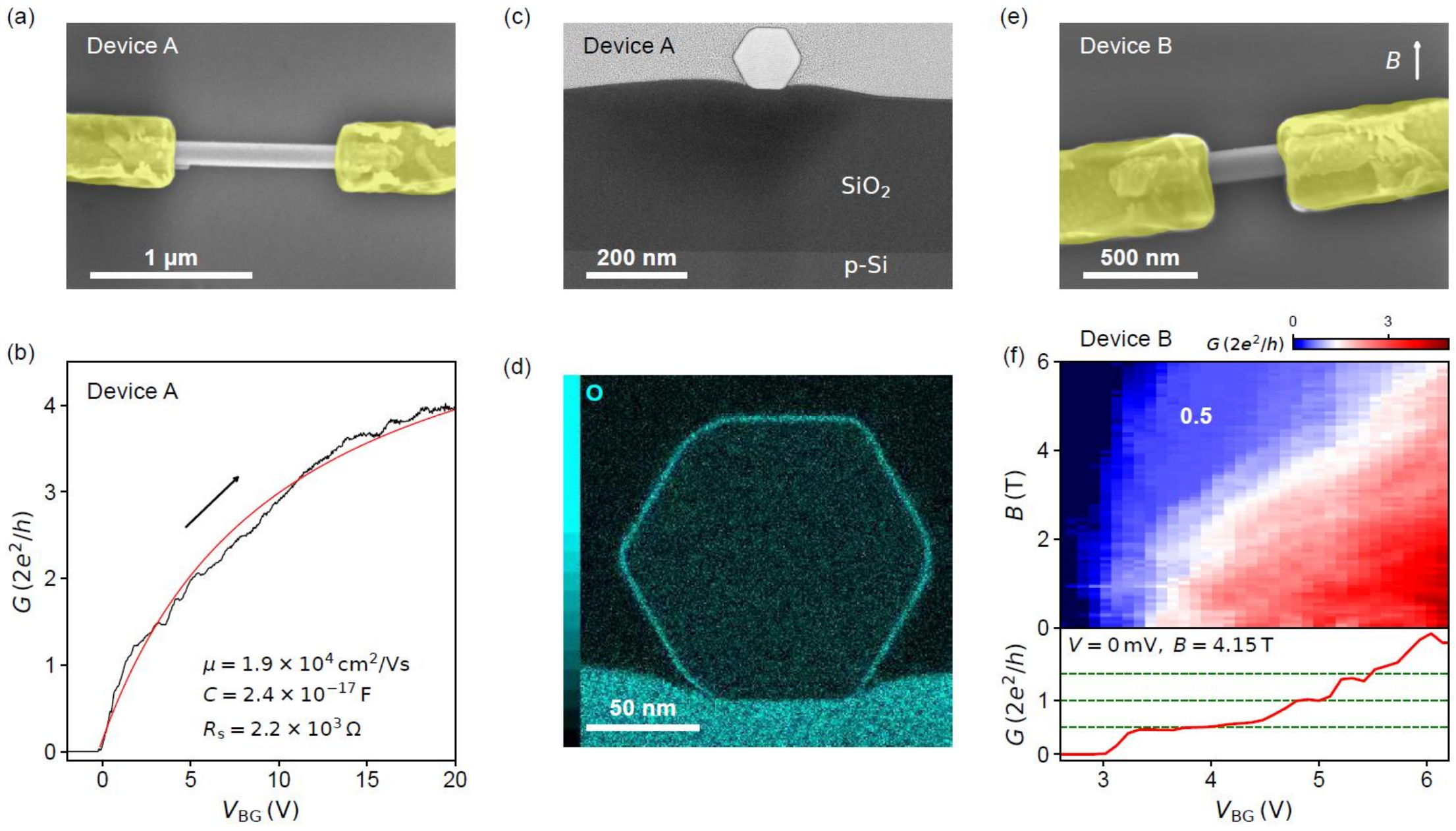

**Figure 5.** Transport characterizations of InSb nanowires. (a) SEM of device A. (b) Pinch-off curve of device A and its mobility fit (red). (c) Cross-sectional STEM of device A. (d) EDS map highlighting the O element. (e) SEM of device B. (f) $G$ as a function of $B$ and $V_{BG}$. $V = 0$ mV. Lower panel, a line cut at 4.15 T.